\pdfoutput=1
\documentclass[aip,jcp,reprint]{revtex4-1}
\usepackage{graphicx}
\graphicspath{{./figs/}}
\usepackage{amsmath} 
\usepackage{notes2bib}
\bibnotesetup{
note-name = ,
use-sort-key = false
}
\usepackage[mathlines]{lineno}% Enable numbering of text and display math
\begin{document}

\title{Johari-Goldstein Relaxation Events Are Metabasin Transitions}
\author{Marcus T. Cicerone}
\email[]{cicerone@nist.gov}
%\homepage[]{Your web page}
%\thanks{}
\affiliation{National Institute of Standards and Technology, Gaithersburg, MD 20899-8543}
\affiliation{Institute for Physical Sciences and Technology, University of Maryland, College Park, MD 20742-2431}
\author{Madhusudan Tyagi}
\affiliation{National Institute of Standards and Technology, Gaithersburg, MD 20899}

\date{\today}

\begin{abstract}
We show that by representing quasi-elastic and inelastic neutron scattering from propylene carbonate (PC) with an explicitly heterogeneous model, we recover signatures of two distinct localized modes in addition to diffusive motion. The intermediate scattering function provides access to the time-dependence of these two localized dynamic processes, and they appear to correspond to transitions between inherent states and between metabasins on a potential energy landscape. By fitting the full q-dependence of inelastic scattering, we confirm that that the Johari-Goldstein ($\beta_{JG}$) relaxation in PC is indistinguishable from metabasin transitions. 
\end{abstract}

% insert suggested PACS numbers in braces on next line
% \pacs{64.70.pm,  66.10.C-,66.30.hh,64.70.ph}

%\maketitle must follow title, authors, abstract, \pacs, and \keywords
\maketitle

% body of paper here - Use proper section commands
% References should be done using the \cite, \ref, and \label commands

\section{Introduction}

It has become clear in the past several decades that dynamic heterogeneity (DH) underlies the characteristic behavior of transport and relaxation processes in glasses and supercooled liquids. The first experimental confirmations of this for glass-forming systems at low temperatures came in the 1990s and were focused on timescales of milliseconds and longer.\citep{SchmidtRohr:PhysicalReviewLetters:1991,Cicerone:Macromolecules:1995,Cicerone:JournalOfChemicalPhysics:1995, Bohmer:EurophysicsLetters:1996} Despite extensive study, only a few general properties of long-time DH have been established, such as approximate lengthscale \citep{Tracht:PhysicalReviewLetters:1998,Reinsberg:JNCS:2002,Chua:ColloidAndPolymerScience:2014,casalini2015dynamic} and lifetime.\citep{Cicerone:JournalOfChemicalPhysics:1995,Bohmer:EurophysicsLetters:1996,Bohmer1998} By contrast, evidence for DH at much shorter times was found the 1980s, \citep{StillingerPhysRevA1983,Miyagawa:TheJournalOfChemicalPhysics:1988} and, because short time DH was observed in simulation, it could be characterization in much better detail. 

Simulation indicates that dynamic heterogeneity at the shortest times appears in the form of intermittent localized molecular rearrangements. Building on the potential energy landscape (PEL) concept,\citep{Goldstein:TheJournalOfChemicalPhysics:1969} Stillinger described these discrete rearrangement events\citep{StillingerPhysRevA1983} in terms of barrier crossings on a high-dimensional PEL with shallow minima corresponding inherent structures (IS), which decorate deeper minima, referred to as metabasins (MB).\citep{debenedetti2001supercooled} It is suggested that transitions between ISs within an MB are associated with local cage distortions, while transitions between MBs appear to involve collective rearrangements of a small number of particles.\citep{Denny:PRL:2003} MB transitions are spatially heterogeneous relaxation events,\citep{Vogel:TheJournalOfChemicalPhysics:2004} and thus appear to be the fundamental element of short-time DH.

Connections have been firmly established between these microscopic collective motions involving particle rearrangements and macroscopic relaxation processes, including self diffusion\citep{Doliwa1:PRE:2003,Starr:JCP:2013,Charbonneau:PNAS:2014,Cicerone:PRL:2014,Cicerone:JNCS:2014} and $\alpha$ relaxation.\citep{Starr:JCP:2013} A connection to the Johari-Goldstein\citep{Johari:TheJournalOfChemicalPhysics:1970} ($\beta_{JG}$) relaxation process has also become increasingly appreciated over the past decade or so. The precise nature of this relaxation process, however, remains an outstanding problem.

Stillinger suggested that $\beta_{JG}$ relaxations correspond to IS transitions in the PEL, with sequential $\beta_{JG}$ relaxations leading to MB transitions and $\alpha$ relaxation.\citep{stillinger1995topographic} However, Vogel et al. have alternatively proposed that exploration of the MB (i.e., a series of IS transitions) should be associated with $\beta_{JG}$ relaxation.\citep{Vogel:TheJournalOfChemicalPhysics:2004} We note here, however that there are many rather remarkable similarities between $\beta_{JG}$ relaxation and the collective relaxation characteristic of MB transitions. Some of these are: i) The $\beta_{JG}$ process appears to bifurcate from the $\alpha$ relaxation when relaxation times are approximately 1 ns, and just when thermal energies are comparable to the heights of potential barriers,\citep{Goldstein:TheJournalOfChemicalPhysics:1969} and when ergodicity times begin to increase substantially.\citep{Thirumalai:PRE:1993} Incidentally, this is the same point at which one expects that the localized molecular reorganization events will become discrete and intermittent rather than occurring continuously.\citep{stillinger1984packing} ii) Rapid rotational jumps of ($6^\circ$ to $10^\circ$) are found for $\beta_{JG}$ relaxation above $T_g$,\citep{Chang:JournalOfNonCrystallineSolids:1994,Vogel:JPhysChemB:2000} corresponding to spatial excursions of $0.2\, r_H$ (assuming that the Stokes-Einstein relation holds locally), and Vogel et al.\citep{Vogel:TheJournalOfChemicalPhysics:2004} found spatial excursions $0.2\,r_H$ to be characteristic of MB transitions. iii) The dielectric $\beta_{JG}$ loss peak exhibits a thermal hysteresis that can be modeled as a relaxation between basins in an asymmetric double well potential,\citep{Dyre:PRL:2003} similar to the localized transitions between local minima in the PEL. iv) It is clear that excursions associated with $\beta_{JG}$ relaxation must not occur as single steps, but as a rapid series of smaller steps\citep{Vogel:JNCS:2002} This is consistent with collective rearrangements\citep{Miyagawa:TheJournalOfChemicalPhysics:1988} characteristic of MB transitions. v) The temperature dependencies of the peak time ($t^\star$) in the non-Gaussian parameter, characterizing short time DH\citep{Starr:JCP:2013}, the mean waiting time between MB transitions\citep{Doliwa1:PRE:2003} and the temperature dependence of the $\beta_{JG}$ relaxation time ($\tau_{\beta,JG}$)\citep{Yu:NationalScienceReview:2014} all seem to follow the temperature dependence of $D_T$, at least to temperatures as low as the mode-coupling critical temperature $T_c$.\citep{Doliwa1:PRE:2003}

Neutron scattering is an ideal tool for investigating the detailed motion of liquids on the timescales and lengthscales germane to the $\beta_{JG}$ relaxation. In fact, we show below that direct signatures of IS transitions, MB transitions, and $\beta_{JG}$ relaxation are present in the neutron scattering. From these signatures, we can show quite clearly that $\beta_{JG}$ relaxation is to be identified with MB transitions. Although neutron scattering has been applied to liquids and glasses for decades, these signatures have not been identified until now. This is probably because, exceptions notwithstanding,\citep{Russina2000,Cicerone:PRL:2014,Vispa:PhysChemChemPhys:2015} neutron scattering from molecular liquids has historically been analyzed in terms of homogeneous models in spite of overwhelming evidence for short-time DH in these systems. 

\section{Results}
\subsection{Frequency Domain Quasielastic Neutron Scattering}

%% FREQUENCY DOMAIN REPRESENTATION  ----------------------------------------------------

\begin{figure}[htbp]%label{SQWwFit}
\begin{center}
\includegraphics[width=9 cm]{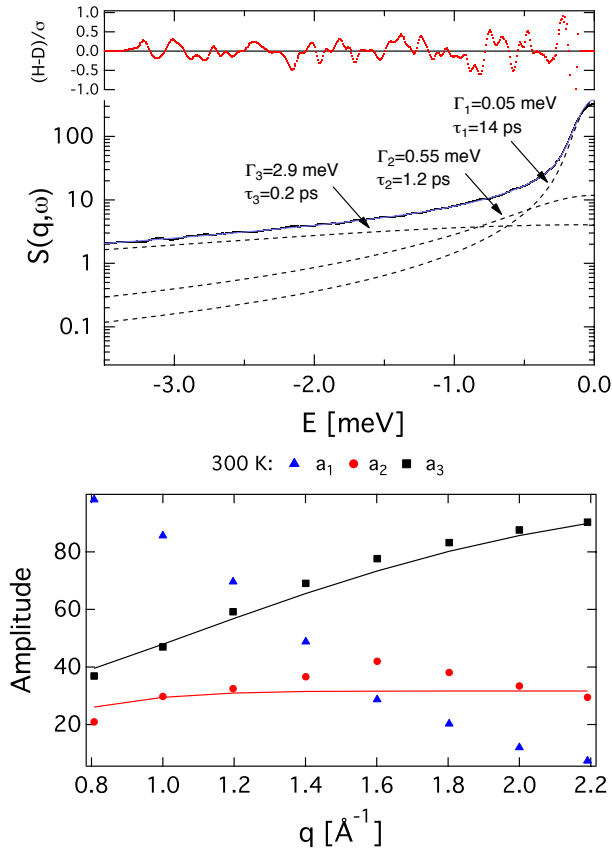}
\caption{Fits to $S(Q,E)$ of PC, measured at T=300 K, using Eq. (\ref{AprxSQE}) Top: Solid black line - $S(q=0.8\AA^{-1},E)$ and fit. The dashed lines are the fit components from three Lorenzians with $\Gamma$ values indicated. Fit residuals are shown in the upper part of this panel. Bottom: Amplitudes of Lorentzian components from fits. \label{SQWwFit}}
\end{center}
\end{figure}

Figure \ref{SQWwFit} shows quasielastic neutron scattering (QENS) results from propylene carbonate (PC). The top panel of Fig. \ref{SQWwFit} shows an example of $S(q,E)$ for PC at 300 K. The data were collected at the NIST neutron center on NG4 with neutron wavelength $\lambda \approx$ 4.0 $\AA$, q in the range (0.22 to 2.77) $\AA^{-1}$, and energy resolution of 200 $\mu$eV.\cite{Copley:ChemicalPhysics:2003} Background and scattering due to methyl rotor motion\citep{Frick:Macromolecules:1994} were accounted for (see supplementary material). The data were binned into 11 discrete q values for analysis.  

Vispa et al. \citep{Vispa:PhysChemChemPhys:2015} recently found that a triple Lorentzian function provided significantly better fits for $S(q,E)$ of a molecular liquid than common models of similar complexity containing functional forms such as KWW and Gaussian. In accordance with their finding, we observe three exponential relaxation processes for PC in time-domain optical Kerr effect data covering similar time and lengthscales (manuscript in preparation). We therefore fit our $S(q,E)$ data with a three-Lorentzian model:
\begin{equation}
S(q,E)=\sum_{i=1}^3a_i(q)\frac{\Gamma_i}{\pi(E^2+\Gamma_i^2)}\label{AprxSQE}
\end{equation}
In order to fit to the data, the model is convolved with a Gaussian function representing the instrument resolution, which is estimated from sample scattering at 30 K. We used an iterative simulated annealing algorithm to find optimized fit parameters, $a_i$ and $\Gamma_i$ at each average q value. For all data reported, the fit residuals were randomly distributed, with amplitudes less than the uncertainty in the data, as exemplified in the top panel of Fig. \ref{SQWwFit}.

The q-dependencies of the fit parameters are shown in the bottom panel of Fig. \ref{SQWwFit}. The low-q drop in intensity of the two broader Lorentzians indicate that they are associated with localized modes, and this is consistent with the fact that their characteristic frequencies are near or above that of the Debye frequency ($\nu_D \approx 1.4$ THz, assuming v=1200 m/s). Accordingly, we fit the amplitudes of these modes ($i$=2,3) as $a_i(q)=c_i\{1-exp[-(\pi\sigma_i q)^2]\}$, assuming Gaussian distributions of displacements, \citep{Rahman:PhysRev:1962} each having distinct characteristic lengthscales. Notably, the two lengtscales obtained in the fit are $\tilde{\sigma}_2=0.2$ and $\tilde{\sigma}_3=0.08$, where $\tilde{\sigma}=\sigma / r_H$, and $r_H=2.6\AA$ is the high-temperature hydrodynamic radius of PC.\citep{Qi:JCP:2000} These values correspond well to the relative average excursions for Lennard Jones particles undergoing MB and IS and transitions respectively.\citep{Vogel:TheJournalOfChemicalPhysics:2004}

\begin{figure}[htbp]%label{GamVsQ2}
\begin{center}
\includegraphics[width=9 cm]{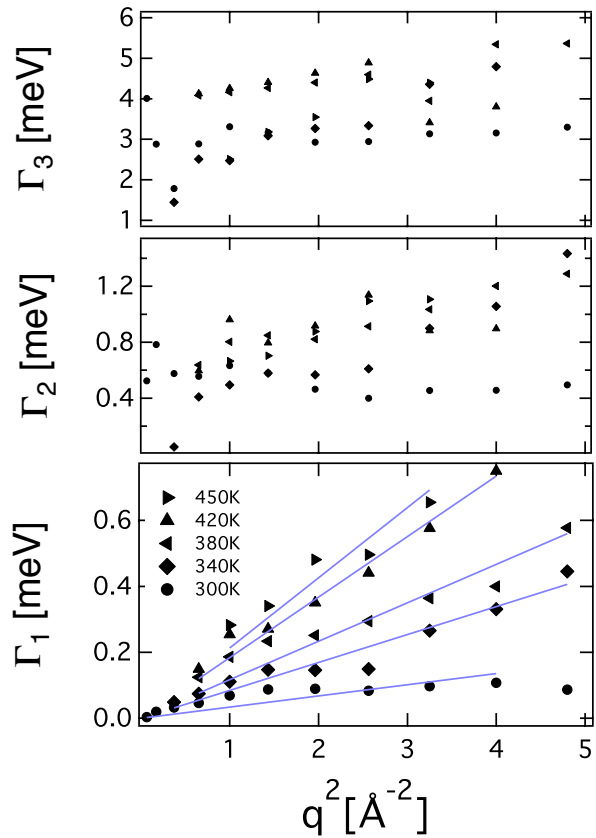}
\caption{Dispersion relations for each of the processes detected. $\Gamma$ values as a function of q from fits to $S(Q,E)$ of PC, at T=300 K, using Eq. (\ref{AprxSQE}). Straight lines in the bottom panel are fits to the data.\label{GamVsQ2}}
\end{center}
\end{figure}

The low-q amplitude rise of the most narrow Lorenzian component ($a_1$) in the lower panel of Figure \ref{SQWwFit} suggests that it is associated with diffusion. This is born out by the $q^2$ dependence of $\Gamma_1$ and the strong temperature dependence for this relaxation process, as shown in the bottom panel of Figure \ref{GamVsQ2}. By contrast, the $\Gamma$ values for the two faster processes are only weakly temperature dependent, indicating a small activation energy, as expected for a highly local motion. Similar behavior of local modes has been noted for other liquids. \citep{FernandezAlonso:JChemPhys:2007,Vispa:PhysChemChemPhys:2015} The faster, 0.2 ps process has been previously associated with overdamped vibrations - rapid, localized collisions between neighboring molecules occurring homogeneously throughout the sample.\citep{FernandezAlonso:JChemPhys:2007} The intermediate process at $\approx 1$ ps has been attributed elsewhere to spatially heterogeneous dynamics in the form of collective molecular rearrangements,\citep{Russina2000,Cicerone:PRL:2014} which is consistent with its assignment to MB transitions. While one could tentatively assign both processes to homogeneous dynamics, we will demonstrate below that a heterogeneous dynamics interpretation is reasonable, whereas a purely homogeneous dynamics interpretation is not. 

Under the assumption that the heterogeneous dynamics assignment is correct, we propose the following model for $S(q,E)$:
\begin{align}
S(q,E)=&(1-\Phi)L_D\otimes[(1-a_v)\,\delta(E)+a_v L_v]\nonumber\\
&+\Phi L_D\otimes[(1-a_v)\,\delta(E)+a_v L_v]\nonumber\\
&\otimes[(1-a_h)\,\delta(E)+a_h L_{h}]\label{FullSQE}
\end{align}
where the terms $a_i$ have the same functional form as in Eq. (\ref{AprxSQE}), $L$ are Lorenzian functions, $\otimes$ is the convolution operator, and the convolutions are over frequency (energy). The two terms in this expression account for two dynamically different classes of molecules; those that can participate in collective rearrangements, and those that cannot. All molecules undergo both diffusion and over-damped vibrations. Accordingly, both terms in Eq. (\ref{FullSQE}) include diffusive ($D$) and vibrational ($v$) components. For a given time window, some fraction ($\Phi$) of molecules can also execute collective rearrangements (hopping) motion ($h$). These are accounted for in the second term of Eq. (\ref{FullSQE}).

When expanded, the expression in Eq. (\ref{FullSQE}) contains four Lorentzian terms, although we indicated above that we needed only three Lorentzians for acceptable fits. In practice, there is no inconsistency here, since two of the Lorentzian terms in the expansion of Eq. (\ref{FullSQE}) are essentially degenerate. Within the model assumption, $L_1$ is due solely to diffusion ($L_1=L_D$), $L_2$ is due to hopping, but slightly broadened by diffusive motion ($L_2 = L_h \otimes L_D$). $L_3$ is due primarily to vibrations, but is a sum of two terms ($L_3$ = $L_v \otimes L_D + L_v \otimes L_h\otimes L_D$), that differ in width by \textless  10\%, and are indistinguishable at the present signal-to-noise ratio. 

\subsection{Time Domain Quasielastic Neutron Scattering}
%% TIME DOMAIN REPRESENTATION -----------------------------------------------------------

In a previous report we showed that time-domain analysis of QENS data at 1 and 10 ps supported a proposed model for liquid relaxation.\citep{Cicerone:PRL:2014} In that work we used a heuristic approximation for the intermediate scattering function, $F(q,t)$. In the present work we analyze a more complete set of time-dependent data analyzing it with a more complete model of $F(q,t)$. Transforming Eq. (\ref{FullSQE}) to the time domain we obtain:

\begin{align}
F(q,t)=&e^{-t\Gamma_D/\hbar}+a_v[e^{-t(\Gamma_D+\Gamma_v)/\hbar}-e^{-t\Gamma_D/\hbar}]\nonumber\\
&+\Phi a_h[e^{-t(\Gamma_D+\Gamma_h)/\hbar}-e^{-t\Gamma_D/\hbar}]\nonumber\\
&+\Phi a_h a_v[e^{-t\Gamma_D/\hbar}-e^{-t(\Gamma_D+\Gamma_h)/\hbar}\nonumber\\
&-e^{-t(\Gamma_D+\Gamma_v)/\hbar}+e^{-t(\Gamma_D+\Gamma_v+\Gamma_h)/\hbar}]\label{FullFqt}
\end{align}
where 
\begin{equation}
\begin{array}{l}
a_i=1-e^{-(\pi q \sigma_i)^2}\,\mbox{(for \textit{i= v, h})}\end{array}\nonumber
\end{equation}
and $\Gamma_D=D_Tq^2$. In fitting the time domain data we fix $\Gamma$ values to those obtained from the frequency domain fits, and allow $\Phi$ and $\sigma$ values to vary. In the regime $t\Gamma_D \ll\hbar \ll t\Gamma_h\leq t\Gamma_v$, we can ignore terms involving $\Gamma_v$ and $\Gamma_h$, and Eq. (\ref{FullFqt}) reduces essentially to
\begin{equation}
F(q)=(1-\Phi)e^{-(q\,\pi\,\sigma_{v})^2}+\Phi e^{-(q\,\pi\,\sigma_{h})^2} 
\label{F2Gauss}
\end{equation}
which we had previously used to fit QENS on several liquids, including PC,\citep{Cicerone:PRL:2014} but using different notation.\bibnote{When approximating Eq. (\ref{F2Gauss}) from a limiting case of Eq. (\ref{FullFqt}), we realized that subscripts $v$ and $h$ were more appropriate than $TC$ and $LC$ as we had previously used.} Both equations yield similar values for fit parameters, but we use the full model here. 

\begin{figure}[htbp]%label{Fqt}
\includegraphics[width=8 cm]{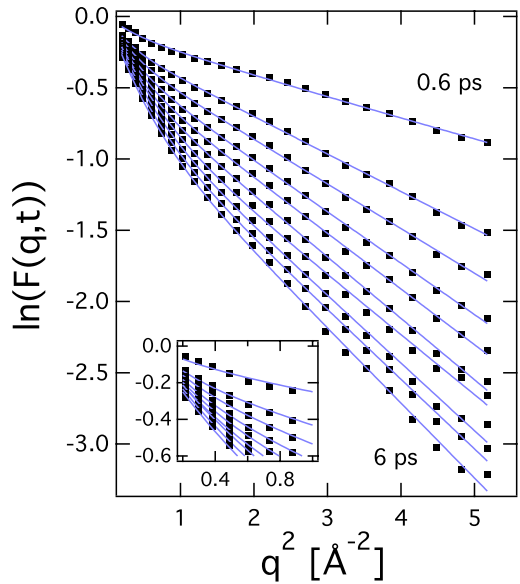}
\caption{$F(q,t)$ for PC at 300 K and times ranging from 0.6 ps to 6 ps in increments of 0.6 ps. Solid lines are fits to Eq. (\ref{FullFqt}). Inset shows expanded low q region. The uncertainties in the data are approximately the size of the symbols. Here, and throughout this paper, error bars indicate uncertainties in parameters at one standard deviation.\label{Fqt}}
\end{figure}

Figure \ref{Fqt} shows $F(q,t)$, transformed to the time domain from $S(q,E)$ for PC at 300 K. The solid lines are fits to the data using Eq. (\ref{FullFqt}). The inset highlights the data and fits at small q, which are difficult to see in the main figure. It is clear that the relaxation is non-Gaussian, as a Gaussian response would result in a straight line on this plot. In fact, there appears to be two linear regimes, corresponding to dynamics on two fairly well defined lengthscales, consistent with the two localized modes inferred from fits to $S(q,E)$. The time-dependence of the lengthscales that we obtain from fits to $F(q,t)$ give us our first strong evidence for and insight into the nature of the heterogeneous dynamics. 

%% TIME EVOLUTION OF SIGMA VALUES ----------------------------------------------------

\begin{figure}[htbp]%label{SigVt}
\includegraphics[width=9 cm]{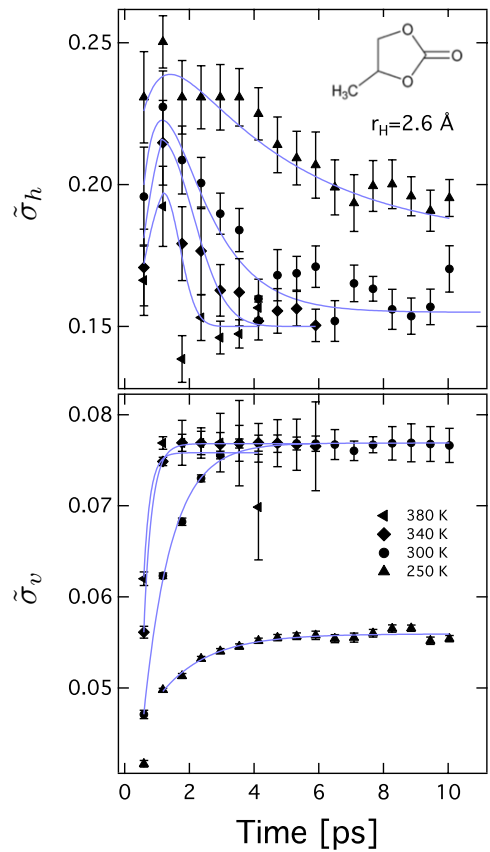}
\caption{Time dependence of $\tilde{\sigma}_{v}$ and $\tilde{\sigma}_{h}$ parameters derived from fits of Eq. (\ref{FullFqt}) to $F(q,t)$ data in the range (250 to 380) K. The solid lines are fits to the data for the 250 K and 300 K, and guides to the eye for the higher temperatures. \label{SigVt}}
\end{figure}

Figure \ref{SigVt} shows $\tilde{\sigma}_v$ and $\tilde{\sigma}_h$ as a function of time for the temperatures indicated. The solid lines are fits to the data for the two lower temperatures, and guides to the eye for the higher temperatures. At all temperatures we observe a non-monotonic time evolution of $\tilde{\sigma}_h$. The initial rise time for $\tilde{\sigma}_h$ is difficult to determine due to limits in the energy range of the instrument, but appears to be $\approx 0.65$ ps, and seems to be insensitive to temperature. On the other hand, the timescale for the subsequent relaxation clearly depends on temperature, being $\approx 0.15 \,\tau_\alpha$ over the range that we can measure it. The timescale for the rise of $\tilde{\sigma}_v$ in the lower panel is also $\approx 0.15 \,\tau_\alpha$.

Many aspectes of the behavior of $\tilde{\sigma}_v$ suggests that it is related to IS transitions. The overall magnitude of the asymtotic values\cite{Vogel:TheJournalOfChemicalPhysics:2004} and the drop in these values beginning somewhat above the mode-coupling critical temperature\citep{HernandezRojas:JNCS:2004} are consistent with that reported for IS transitions. Also, the rapid rise to a plateau suggests \citep{Middleton:PRB:2001} exploration of a bounded phase space, such as a series of IS transitions within a single MB. Whether $\tilde{\sigma}_v$ is characteristic of a small number of IS transitions or many of them could be decided if an average IS transition rate were known. The expected IS transition rate is, however, difficult to estimate from published simulation studies due to the computational expense of checking for these transitions.\citep{Doliwa1:PRE:2003} There are two limiting scenarios. If IS transition rates are $\approx$ 1 THz, $\tilde{\sigma}_v$ values would represent a very small number of IS transitions. The data would then suggest that larger displacements take slightly longer to occur due to displacement-dependent energy barriers. On the other hand, if IS transitions are much faster, $\tilde{\sigma}_v$ would represent many IS transitions, and provide a metric for the characteristic displacement of molecules during exploration of a MB. The primary difference between these two scenarios is the IS barrier height relative to thermal energy, and the dominant scenario may change with temperature. At the temperatures for which we plot $\tilde{\sigma}_v(t)$, it would seem that the rapid transition scenario is most likely for at least two reasons. One is that barriers for IS transitions are significantly smaller than those of MB transitions,\citep{Middleton:PRB:2001} and we shal see below that the latter occur on the order of 1 ps at high temperatures. The other is that we observe no evidence of reversing transitions, which are known to occur for IS transitions.\citep{Vogel:TheJournalOfChemicalPhysics:2004} We suggest that the absence of an obvious signature for the reversing transitions is that individual IS transitions occur at a rate that is out of our experimental window (i.e., \textgreater 2 THz). 

We note also that we cannot completely separate effects of IS transitions from effects of damped vibrations. We suggest that the latter would reach a plateau value after only a few collisions (\textless 100 fs), whereas exploration of a MB through successive IS transitions could lead to $\tilde{\sigma}_v$ values that increase over several ps as observed.

We find that $\tilde{\sigma}_h$ appears to be associated with MB transitions. The relative amplitudes we observe are similar to that reported for mean displacement during MB transitions for a Lennard-Jones system.\cite{Vogel:TheJournalOfChemicalPhysics:2004} Further, the non-monotonic behavior was previously shown to arise from molecular hopping associated with collective rearrangements\citep{Cicerone:JNCS:2014} where we observed a fraction of molecules making large excursions in a small number of steps, with some of them subsequently returning to their origin. The timescale for the reversing rearrangements appears to be the same as the rise time of $\tilde{\sigma}_v$, suggesting that this is indeed the the time required to explore MBs, as suggested above. The measured rise time ($\approx 0.65$ ps) is likely characteristic of the barrier crossing time for the transitions. The wait time between MB transitions has been linked to diffusion\citep{Doliwa1:PRE:2003}, and is of particular interest in connection with molecular relaxation. Information on the wait times between MB transitions is contained in $\Phi(t)$.

%% INTRODUCE KINETIC EQUATIONS -----------------------------------------------------------------

We next investigate the time dependence of the fitting parameter $\Phi$. Before doing so, we clarify the meaning of this parameter, and this is best done through appeal to a simple model. $\sigma_v$ and $\sigma_h$ represent two distinct types of motion, exploration of MBs through IS transitions, and transitions between MBs respectively. IS transitions should be executed continuously by all molecules, whereas MB transitions will be executed by only a subset of molecules at any given time.\citep{Vogel:TheJournalOfChemicalPhysics:2004} We thus consider a minimal model with two dynamic states where molecules can undergo collective rearrangements (MB transitions) in one state and not in the other. Given the emerging connection between local structure and dynamics,\citep{Kawasaki:PhysicalReviewLetters:2007,Leocmach:NatComm:2012} we assume that molecules which are locally more highly ordered, or tightly caged (TC) by their neighbors have barriers between MBs that are too high to overcome, and these cannot execute collective motion. Likewise, we assume that molecules in slightly less ordered regions, molecules are more loosely caged (LC), and have smaller barriers between MBs allowing them to executing collective rearrangements. The assumed association between local ordering and ability to execute collective motion is not critical to, but is convenient for the discussion that follows.

We assume a dynamic equilibrium between TC and LC states:
\begin{eqnarray}
\mbox{}&k_{TL}&\mbox{}\nonumber\\
TC &\rightleftharpoons & LC\\
\mbox{}&k_{LT}&\mbox{}\nonumber
\end{eqnarray}
where, by detailed balance,
\begin{equation}
\frac{n_{LC}}{n_{TC}}=\frac{k_{TL}}{k_{LT}}
\end{equation}
and, we define
\begin{equation}
\Phi_0\equiv\frac{n_{LC}}{n_{LC}+n_{TC}}
\end{equation} 
where the $n_{LC}$ and $\Phi_0$ are the instantaneous number and fraction of molecules in LC domains respectively. 

We showed previously that the earliest time at which TC and LC states are clearly distinguished is approximately 1 ps, and that this value is independent of temperature.\citep{Cicerone:JNCS:2014} We detect TC or LC states only through the statistical properties of their displacements. Differences in these displacements develop as molecules explore the cage formed by their neighbors, and $\approx$ 1 ps is required for this process.\citep{ottochian:PhilMag:2010} Accordingly, we set $\Phi_0=\Phi(1.1\,ps)$ from fits of QENS data to Eq. (\ref{FullFqt}).

%% PRESENT \PHI(T) -----------------------------------------------------------------
Fits to the neutron scattering measurements provide us with $\Phi(t)$, the fraction of  molecules that have executed a large displacement (i.e., participated in an MB transition) up to time t. Thus, for $t\, \textgreater$ 1 ps, an expression for $\Phi(t)$ can be written as:
\begin{equation}
\Phi(t)=\Phi_0+\int_{0}^{t}(1-\Phi(t'))k_{TL}(t') dt' 
\end{equation}
For $k_{TL}=\tau_{TL}^{-1}=$ constant, $\Phi(t)=\Phi_0+(1-e^{-t/\tau_{TL}})$. 

\begin{figure}[htbp]%label{PhiVt}
\begin{center}
\includegraphics[width=8 cm]{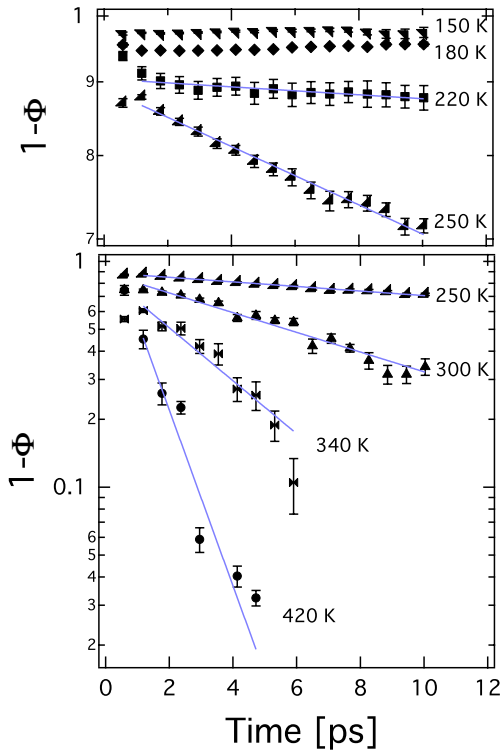}
\caption{Time dependence of $\Phi$ derived from fits of Eq. (\ref{FullFqt}) to $F(q,t)$ for temperatures indicated. Dashed lines are exponential to the data.\label{Phivt}}
\end{center}
\end{figure}

Figure \ref{Phivt} shows the time dependence of $(1-\Phi(t))$ at the temperatures indicated. At the two lowest temperatures shown we see a plateau in $(1-\Phi(t))$. Thus, TC and LC states do not exchange (i.e., LC states doe not propagate) appreciably, and $\Phi(t)\approx\Phi_0$ over the time range of the experiment for these temperatures. On the other hand, for times \textgreater 1 ps, and temperatures $\geq$ 220 K, we observe exponential decrease in $(1-\Phi(t))$, indicating LC states propagate such that all molecules eventually participate in collective rearrangements at these temperatures. 

\begin{figure}[htbp]%label{KTLvT}
\begin{center}
\includegraphics[width=9 cm]{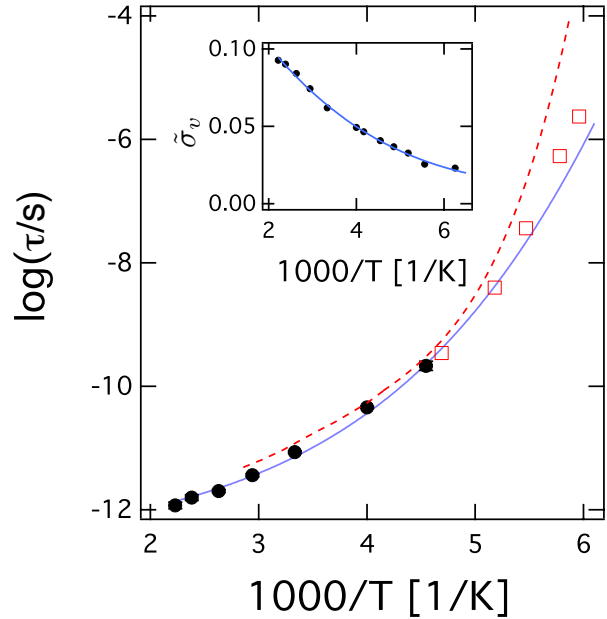}
\caption{PC Relaxation times: (black circles) $\tau_{TL}$ calculated from time dependence of $\Phi$ shown in Fig. \ref{Phivt} error bars drawn are slightly smaller than the size of the symbols.  (solid blue line) fit to $\tau_{TL}$ data using Eq. (\ref{HW}), with $\delta = 1.70\pm.07$ kJ/mole and $\tau_0=0.35\pm0.02$ ps. (open squares) $\beta_{JG}$ relaxation times from dielectric spectroscopy\citep{Ngai:TheJournalOfChemicalPhysics:2001}  and (dashed line) $\alpha$ relaxation times, from dielectric spectroscopy\citep{Ngai:TheJournalOfChemicalPhysics:2001} and light scattering.\citep{du1994light} The dielectric data were reported as peak frequencies, and were shifted to coinside with $\alpha$ relaxation times from light scattering. Inset: $\tilde{\sigma}_{TC}$ values used as input to Eq. (\ref{HW}). The solid line in the inset is a polynomial fit.\label{KTLvT}}
\end{center}
\end{figure}

The $\tau_{TL}$ values obtained from QENS data at $T\geq$ 220 K are plotted as solid circles in Fig. \ref{KTLvT}. At these high temperatures, $\tau_{TL}\approx 0.6 \tau_\alpha$. Values of $\tau_\alpha$ are obtained from dielectric relaxation\citep{Ngai:TheJournalOfChemicalPhysics:2001} and light scattering,\citep{du1994light} and are represented by a dashed line in Fig. \ref{KTLvT}. The high-temperature correspondence between $\tau_{TL}$ and $\tau_\alpha$ suggests that transitions between MBs plays an important role in $\alpha$ relaxation, as previously suggested,\citep{stillinger1995topographic} but the timescales appear not to be identical. We will discuss this relationship in more detail below.

%% Introduce Hall-Wolynes ----------------------------------------------------

We now consider the physical processes associated with propagation of LC domain in order to formulate an expression for $\tau_{TL}$. From the perspective of the PEL framework, once a MB transition occurs involving rearrangement of a small group of molecules, a new set of ISs become active,\citep{Vogel:TheJournalOfChemicalPhysics:2004} and a distinct group of molecules will be involved in the next cooperative rearrangement that signals a transition into the next new MB. Thus, $\tau_{TL}$ is the MB transition rate, and it should be related to the time required to explore the MB by sampling the IS states within it.

The Hall-Wolynes (HW) ansatz\citep{Hall:JCP:1987} seems to be appropriate for estimating propagation rates. In fact, this model is perhaps even more appropriate for modeling a local barrier-crossing process than the many-particle process implicated in $\alpha$ relaxation. HW assumed that potential wells can be approximated as parabolic near the bottom, and the height of the barrier between wells is proportional to the distance between their minima ($\sigma_0$) in configuration space. With this, they determined that the logarithm of the barrier crossing rate should depended on the mean squared particle displacement within the well ($\left <\sigma \right >$)as $(\sigma_0/\left <\sigma \right >)^2$. 

Since we are interested in estimating MB transition rates, the appropriate interbasin distance ($\tilde{\sigma}_0$) is related to $\tilde{\sigma}_h$. The latter value varies only slightly with temperature, and we have argued that these variations are primarily only apparent variations due to kinetic effects.\citep{Cicerone:PRL:2014} Recognizing that $\tilde{\sigma}_h$ drops in time due to reversing transitions, we take $\tilde{\sigma}_0$=0.28, the maximum value measured for $\tilde{\sigma}_h$ (see Fig. \ref{SigsAt1ps}). We can assume that $\tilde{\sigma}_v$ represents the displacement within the MB well. We find that the original HW approach works well at low temperature, where $\partial{\tilde{\sigma}_v}/\partial{T} = const.$, however, at higher temperature, we allow that relationship to vary, and reformulate the HW relationship as:

\begin{equation}
\tau_{TL}=\tau_0 \, exp\left [\frac{\delta\,\tilde{\sigma}_0}{\tilde{\sigma}_{v}kT} \right]
\label{HW}
\end{equation}
 
Figure \ref{KTLvT} shows $\tau_{TL}$ values obtained from the data of Fig. \ref{Phivt}, and a fit to those values using Eq. (\ref{HW}). We obtain $\delta = 1.70\pm.07$ kJ/mol, which is different than we previously reported.\citep{Cicerone:PRL:2014} The present value is obtained from data over a wider temperature range, and is normalized using a different $\tilde{\sigma}_0$.  

We note that the our lowest temperature $\tau_{TL}$ data from QENS seems to match up precisely with the dielectric $\tau_{\beta, JG}$ data, however, it is also not so different from the dielectric $\tau_\alpha$ data, making it difficult to determine which it corresponds best to. On the other hand, the HW fit to $\tau_{TL}$ extrapolates precisely into the $\beta_{JG}$ relaxation times, rather than the $\alpha$ relaxation times. Although suggestive, this is only an extrapolation, and we seek further evidence to determine which, if either, of relaxation process the MB transitions are to be associated with. 

\subsection{Elastic Incoherent Neutron Scattering}
%% HFBS DATA -----------------------------------------------------------------

As we demonstrate below, we can use elastic incoherent neutron scattering (EINS) to estimate $\tau_{TL}$ values at temperatures below the bifurcation point between $\tau_\alpha$ and $\tau_{\beta,JG}$. EINS measurements were performed at the NIST Center for Neutron Research on the High Flux Backscattering (HFBS) spectrometer\citep{gehring1997PhysicaB} with an incident neutron wavelength of 6.271 $\AA$ and a 0.85 $\mu$eV full width at half-maximum energy resolution and a momentum transfer (q) range of (0.25 to 1.35) $\AA^{-1}$. The spectrometer operates in the fixed-window mode where the elastic scattering intensity is recorded as a function of q while the sample is cooled at 1 K/min from 345 K to 4 K. The inelastic scattering is binned into discrete q values, and scattering intensity is integrated over the instrument resolution:

\begin{equation}
I(q,\gamma_R)=\int_{-\gamma_R}^{\gamma_R}S(q,E) dE \label{I(q)}
\end{equation}

\begin{figure}[htbp]%label{HFBS}
\includegraphics[width=9 cm]{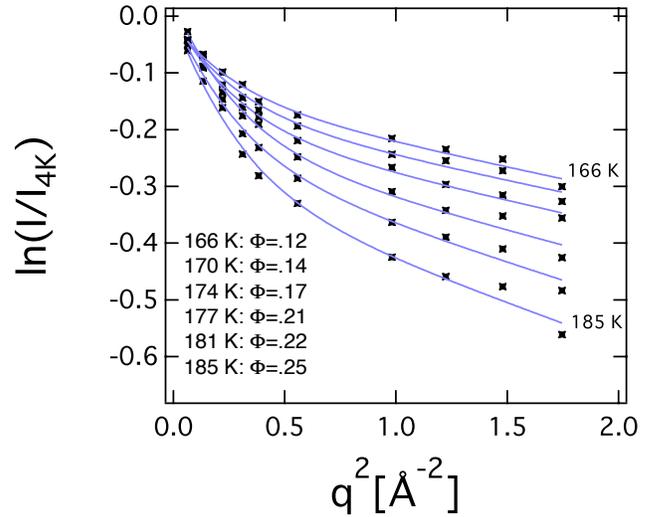}
\caption{Elastic scattering intensity as a function of $q^2$, T = 168, 175, 183, and 190K, referenced against scattering acquired at 4 K. \label{HFBS}}
\end{figure}

Intensity vs q data are shown in Fig. \ref{HFBS} for the temperatures indicated. The data are typically analyzed assuming a harmonic oscillator model (assuming $S(q) \propto \mbox{exp}(-q^2<u^2>)$, where $<u^2>$ is a mean-squared displacement. As is evident, the log(I) vs q plots deviate significantly from linearity at low q, but are approximately linear for higher q values. Thus, the low q data are typically ignored. Here, we use the more complete model in Eq. (\ref{FullSQE}) for $S(q,E)$. The required integration is trivial, as the Lorenzian terms are simply replaced by $2\,\pi^{-1} \mbox{arctan}(\gamma_R/\Gamma_i)$. From this data we obtain fits at only a single time point, $t_R = \hbar/\gamma_R \approx $ 2ns. On this timescale, terms containing $\Gamma_v$ and $\Gamma_h$ are not important, and only variations in $\Phi$, $\Gamma_D$, $\sigma_v$ and $\sigma_h$ impact the data fits.

The HFBS data and fits using Eq.s (\ref{FullSQE}) \& (\ref{I(q)}) are shown in Fig. \ref{HFBS}. Fitting is performed by constraining $\Gamma_D$ and $\sigma_h$ to values obtained from fits to the DCS data, and allowing $\sigma_v$ and $\Phi$ to vary. We obtain values for $\sigma_v$ that are $\approx$ 50\% larger than those obtained at 1 ps for the same temperatures. The $\Phi$ values obtained from these fits are displayed in Fig. \ref{HFBS}. Constraining $\Gamma_D$, $\sigma_h$, and $\sigma_v$ leads to slightly degraded fits, but similar $\Phi$ values.

\begin{figure}[htbp]%label{TauJG}
\begin{center}
\includegraphics[width=9 cm]{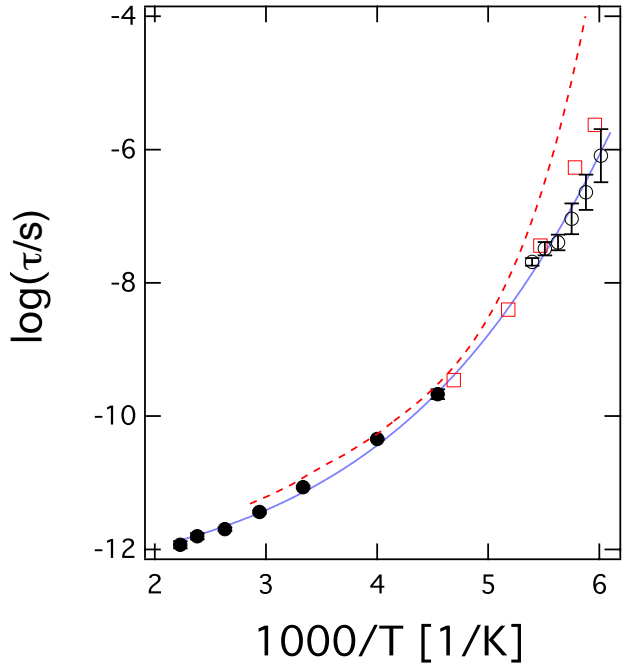}
\caption{(black solid circles) $\tau_{TL}$ calculated from time dependence of $\Phi$ shown in Fig. \ref{Phivt} error bars drawn are slightly smaller than the size of the symbols. (black open circles) $\tau_{TL}$ calculated from fits of Eq. (\ref{PhiHFBS}) to HFBS data. (solid blue line) fit to $\tau_{TL}$ data using Eq. (\ref{HW}), with $\delta = 1.70\pm.07$ kJ/mole and $\tau_0=0.35\pm0.02$ ps. (open squares) $\beta_{JG}$ relaxation times.\label{TauJG}}
\end{center}
\end{figure}

With $\Phi$ values at 1 ps ($\Phi_0$) and at $t_R=$2 ns, we can obtain values for $\tau_{TL}$ if we know the form of the relaxation function for $\Phi(t)$. We wish to test the hypothesis that $\tau_{TL}$ (the rate of MB transitions on the PEL) is related to the $\beta_{JG}$ relaxation process.  Thus we assume a Cole-Cole form, as is found for Johari-Goldstein relaxation. Accordingly, we solve the following equation for $\tau_{TL}$:

\begin{align}
\frac{\Phi(t_R)-\Phi_0}{1-\Phi_0}=&\int_{0}^{\gamma_R} C(\omega,\tau_{TL}) d\omega\bigg/\int_{0}^{\infty} C(\omega,\tau_{TL}) d\omega\label{PhiHFBS}
\end{align}
where $C(\omega,\tau_{TL})$ is the imaginary component of the Cole-Cole distribution
\begin{equation}
C(\omega,\tau_{TL})= Im\left[\frac{1}{1+(i\omega \tau_{TL})^{1-\alpha_{CC}}}\right]\nonumber \\
\end{equation}
and where we have parameterized $\alpha_{CC}$ as a function of temperature from the data of Ngai et al.\citep{Ngai:TheJournalOfChemicalPhysics:2001} as $\alpha_{CC}=-1.697+382/T$. We calculate $\tau_{TL}$ values from EINS over a temperature range limited above by instrument time window, (the point at which $\tau_{TL} \leq 0.1\, t_R$) and below by $T_g$, since our PC sample was quenched relatively quickly. The $\tau_{TL}$ values estimated in this way are plotted in Fig. \ref{TauJG}. These values correspond precisely to $\tau_{\beta,JG}$ values, differing from $\alpha$ relaxation times by more than three orders of magnitude near $T_g$.

We believe this is the first time that $\beta_{JG}$ relaxation has been identified from neutron scattering data, although Sperl previously proposed a Cole-Cole form for suceptabilities that accounted for dynamics observed over a similar length and timescale in optical Kerr effect data.\citep{Sperl:PhysicalReviewE:2006}

\section{Discussion}
\subsection{Dynamic Heterogeneity?}
Throughout this paper we have assumed that the observed motion on two distinct lengscales is due to heterogeneous dynamics. This assumption is not without precedent, as we \citep{Cicerone:PRL:2014} and others, \citep{Vispa:PhysChemChemPhys:2015,Russina2000} have previously presented evidence that the $\approx$ 1 ps response is due to collective, dynamically heterogeneous motion. Furthermore, there is overwhelming evidence for heterogeneous, collective dynamics from simulation.\citep{stillinger1984packing,Vogel:TheJournalOfChemicalPhysics:2004,Keys:PhysicalReviewX:2011}

In spite of significant circumstantial evidence to the contrary, it possible in principle that the distinct lengthscales of dynamics observed here arise from homogeneous dynamics. Thus, we briefly review the evidence for the heterogeneous dynamics case. Strong evidence can be found for short time DH in the time and temperature dependence of the scattering. Quite apart from the particular values of fit parameters, there are several trends in the scattering data that must be accounted for, and are not compatible with homogeneous dynamics.

It is clear from the data in Fig. \ref{Fqt} that more than one lengthscale of motion contributes significantly to $F(q,t)$. A Gaussian q-dependence indicating a single characteristic lengthscale for motion would be represented by a straight line in this figure. By contrast, the data are apparently bi-linear and are fit very well with a simple two-Gaussian model. In a homogeneous model, distinct lengthscales can be explained only through anisotropic or intramolecular motion. Methyl rotor motion is the only possible intramolecular motion for PC, however, the characteristic lengthscale for the methyl rotor would be roughly 2 times smaller than could be detected in the q range used for the QENS experiments. Thus, methyl rotor motion is not responsible for either of the modes of motion that we detect. 

\begin{figure}[htbp]%label{SigsAt1ps}
\includegraphics[width=9 cm]{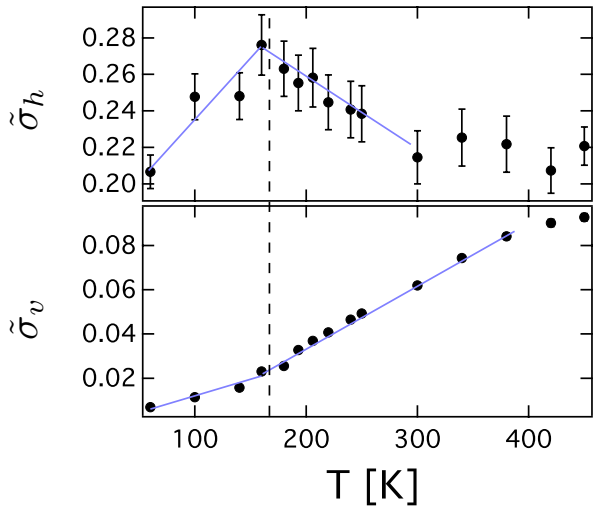}
\caption{$\tilde{\sigma}_v$ and $\tilde{\sigma}_h$ measured at 1 ps as a function of temperature. The solid lines are guides to the eye. The vertical dashed line marks $T_g =156$ K for PC. \label{SigsAt1ps}}
\end{figure}

Anisotropic motion also appears not to be responsible for the two lengthscales of motion we see. Figure \ref{SigsAt1ps} shows the temperature dependence of $\tilde{\sigma}_h$ and $\tilde{\sigma}_v$ values obtained at 1 ps. The $\tilde{\sigma}_h$ values don't vary by more than 50\% over the temperature range explored (60 K to 450 K), whereas the $\tilde{\sigma}_v$ values change by almost a factor of 20 over this same range. The very different temperature dependencies of these two lengthscales could result from anisotropic motion only in the presence of significant temperature-dependent ordering in the liquid, for which there is no evidence. In the absence of such ordering, the ratio of lengthscales would be a function of the molecular geometry, and temperature-independent. 

As with the temperature dependence of the ratio of $\tilde{\sigma}_v$ and $\tilde{\sigma}_h$, we know of no explanation based in homogeneous dynamics for the non-monotonic time-dependence of $\tilde{\sigma}_h$, or the fact that the intensity of the lower-q scattering increases (i.e., $\Phi$ increases) with time. We thus conclude that the scattering signatures are indeed due to collective dynamics that are spatially and temporally heterogeneous.

\subsection{Insights Into $\beta_{JG}$ Relaxation}\label{aandb}

The $\beta_{JG}$ relaxation  was first identified in 1970 for rigid molecular glassformers.\citep{Johari:TheJournalOfChemicalPhysics:1970} Having clear intermolecular origins, it is a collective relaxation process, and seems to emerge in all non-crystalline systems.\citep{Ngai:TheJournalOfChemicalPhysics:2001,Yu:NationalScienceReview:2014} Stillinger proposed that the $\beta_{JG}$ could be related to IS transitions, while the $\alpha$ relaxation process could be associated with MB transitions.\citep{stillinger1995topographic} However, Vogel et al.\citep{Vogel:TheJournalOfChemicalPhysics:2004} later pointed out that IS transitions are essentially single-step events, whereas $\beta_{JG}$ relaxations are multi-step events.\citep{Vogel:JPhysChemB:2000,Vogel:JNCS:2002} They also pointed out that and individual MB transitions did not relax molecules sufficiently to qualify as the $\alpha$ process. Instead, these authors suggested that $\beta_{JG}$ relaxation should be associated with exploration of individual metabasins through multiple IS transitions, and that the $\alpha$ relaxation resulted from a series of MB transitions.\cite{Vogel:TheJournalOfChemicalPhysics:2004}  

Here we have measured the characteristic time of both the exploration time within an MB and transition rates between MBs. We find that the former is approximately $0.15\tau_\alpha$, so about 4 times faster than expected for $\tau_{\beta,JG}$ at the lowest temperatures we could directly measure it. Instead, we find that transitions between MBs, which we detect through large lengthscale cooperative motion, have a characteristic time that is identical to that of $\tau_{\beta,JG}$ from the temperature at which it bifurcates from $\tau_\alpha$ down to the glass transition temperature, $T_g$. 

As we discussed in the introduction, there are many commonalities in the known properties of MB transitions and $\beta_{JG}$ relaxation. Included in these are the assymetric double well behavior of $\beta_{JG}$ relaxation,\citep{Dyre:PRL:2003} the multi-step nature of both $\beta_{JG}$\citep{Vogel:JPhysChemB:2000,Vogel:JNCS:2002} and MB transitions\citep{Middleton:PRB:2001,Doliwa1:PRE:2003,Vogel:TheJournalOfChemicalPhysics:2004}, the correspondence between rotational jump angles for $\beta_{JG}$\citep{Chang:JournalOfNonCrystallineSolids:1994,Vogel:JPhysChemB:2000,Saito:PRL:2012} and jump distances for MB transitions,\citep{Vogel:TheJournalOfChemicalPhysics:2004} and the fact that all molecules appear to participate in $\beta_{JG}$\citep{Vogel:JPhysChemB:2000} relaxation and MB transitions, \citep{Vogel:TheJournalOfChemicalPhysics:2004} although not all at once.

One point of apparent discrepancy between expected behavior of MB transitions and $\beta_{JG}$ is that the strength of the former is determined by $\Phi_0$ and drops monotonically with reduced temperature,\citep{Cicerone:PRL:2014} whereas it was reported that the $\beta_{JG}$ relaxation strength drops with temperature and plateaus at $T_g$.\citep{Vogel:JPhysChemB:2000} Here we clarify this statement: While it appears that the ratio of the $\alpha$ and $\beta_{JG}$ relaxation strengths reaches a plateau near $T_g$, for some rigid glassformers,\citep{Kudlik:JMS:1999} this statement doesn't seem to hold for absolute strength of the $\beta_{JG}$ process. The strength of the $\beta_{JG}$ relaxation is small in the region of $T_g$, so further reduction may escape detection without careful analysis, but it is quite clear that the absolute strength of the $\beta_{JG}$ relaxation continues to drop with reduced temperature even below $T_g$.\citep{Johari:ANYAS:1976,Kudlik:EPL:1997,Kudlik:JMS:1999,Ngai:TheJournalOfChemicalPhysics:2001} Thus, this apparent discrepancy is resolved. 

Our association of MB transitions with $\beta_{JG}$ sheds light on the latter as the PEL framework is now fairly well developed from a configuration space perspective.\citep{Middleton:PRB:2001,Doliwa2:PRE:2003} There has also been some work on the real-space properties of MB transitions.\citep{Middleton:PRB:2001,Vogel:TheJournalOfChemicalPhysics:2004,Keys:PhysicalReviewX:2011,Schoenholz:NatPhys:2016} Of particular note is the work of Middleton et al.\citep{Middleton:PRB:2001} who discuss MB transitions for strong and fragile glassformers. They indicate that many of these rearrangements are quite tractable, and similar to vacancy creation in crystalline solids, although some of the higher energy rearrangements are much more exotic, with higher degrees of cooperativity. 

\section{Conclusions}

In this paper, we propose a model for neutron scattering in amorphous systems that explicitly includes  heterogeneous dynamics. Upon applying the model to quasielastic neutron scattering (QENS) data of propylene carbonate, we revealed motion that corresponds to exploration of metabasins (MBs) through inherent state (IS) transitions, and to MB transitions on a potential energy landscape (PEL). In spite of more than 50 years of theoretical and simulation work on PELs, this is the first time to our knowledge that these classes of transitions have been identified in experimental data. Further, upon applying the model to incoherent elastic neutron scattering (IENS) data of PC, we were able to show that the characteristic time for MB transitions is identical to that of the Johari-Goldstein ($\beta_{JG}$) relaxation. 

\begin{acknowledgments}
Official contributions of the National Institute of Standards and Technology. Not subject to copyright in the United States.
\end{acknowledgments}

% Create the reference section using BibTeX:
\bibliography{TwoState}

\begin{thebibliography}{54}
\bibitem [{1()}]{1}%
  \BibitemOpen
  \href@noop {} {}\bibinfo {note} {When approximating Eq. (\ref {F2Gauss}) from
  a limiting case of Eq. (\ref {FullFqt}), we realized that subscripts $v$ and
  $h$ were more appropriate than $TC$ and $LC$ as we had previously
  used.}\BibitemShut {Stop}%
\end{thebibliography}
\end{document}